\documentclass[
 twocolumn,
 amsmath,amssymb,
 aps, 
 physrev,
 10pt,
 floatfix,
superscriptaddress,
nobibnotes,
]{revtex4-2}
\usepackage{graphicx}% Include figure files
\usepackage{dcolumn}% Align table columns on decimal point
\usepackage{bm}% bold math
\usepackage{amsmath}
\usepackage{amssymb}
\usepackage{xcolor}
\usepackage{comment}
\usepackage{hyperref}% add hypertext capabilities
\usepackage[normalem]{ulem}
\usepackage{physics}
\usepackage{tabularx}
\newcommand{\floor}[1]{\left\lfloor #1 \right\rfloor}

\begin{document}

\title{Optimized Spectral Purity of Heralded Single Photons at the Telecom O-Band}

\author{Wu-Hao Cai}
\email{cai.wuhao.t7@dc.tohoku.ac.jp}
\affiliation{Graduate School of Science, Tohoku University, 6-3, Aramaki Aza-Aoba, Sendai, 980-8578, Japan}

\author{Soyoung Baek}
%\email{baek.soyoung.e3@tohoku.ac.jp}
\affiliation{Graduate School of Science, Tohoku University, 6-3, Aramaki Aza-Aoba, Sendai, 980-8578, Japan}

\author{Rui-Bo Jin}
%\email{jrb@hunnu.edu.cn}
\affiliation{Key Laboratory of Low-Dimensional Quantum Structures and Quantum Control of Ministry of Education, Department of Physics and Synergetic Innovation Center for Quantum Effects and Applications, Hunan Normal University, Changsha 410081, China}

\author{Fumihiro Kaneda}
\email{fumihiro.kaneda.b6@tohoku.ac.jp}
\affiliation{Graduate School of Science, Tohoku University, 6-3, Aramaki Aza-Aoba, Sendai, 980-8578, Japan}
\affiliation{PRESTO, Japan Science and Technology Agency, Kawaguchi, 332-0012, Japan}

\date{\today}% It is always \today, today,

\begin{abstract}
We report on optimizing the spectral purity of heralded single photons in the telecom O-band, where single photons can be propagated with low loss and low dispersion in a standard telecom optical fiber. 
We numerically searched for various group-velocity-matching conditions and corresponding optimal poling structures of a potassium titanyl phosphate crystal for spontaneous parametric downconversion.
Our poling optimization results using phase-matching coherence-length and sub-coherence-length modulation schemes show $> 99.4$\% spectral purity with pump wavelengths ranging from 603.8 nm to 887.3 nm. Some optimized configurations are feasible with off-the-shelf lasers and single-photon detectors. 
Moreover, by investigating noise photon spectra for different poling optimization methods, we show that, in practice, appropriate, gentle spectral filtering helps achieve high purity. 
This study will pave the way for developing practical quantum sources for quantum information applications at the telecom O-band. 
\end{abstract}

\maketitle

%\tableofcontents

\section{Introduction}

The telecom O-band, a wavelength range from 1260 nm to 1360 nm, is advantageous in the low loss and the presence of the zero-dispersion wavelength (1310 nm) in standard telecom optical fiber cables. Those features allow for more efficient channel allocation, next-generation Ethernet \cite{Chorchos2019}, and larger capacity coherent transmission compared to the telecom C-band \cite{Berikaa2024, Elson2024}. Additionally, the O-band is used for upstream wavelengths in passive optical network systems \cite{Abbas2016}. The telecom O-band also offers superior performance in optical coherence tomography imaging in the medical field compared to the C-band \cite{Biedermann2009, Potsaid2012}. 

While those applications utilize intense classical light sources, the low-loss and low-dispersion characteristics can also benefit photonic quantum information applications.
A key quantum effect in photonic quantum information processing is two-photon interference by which qubit-qubit quantum gates \cite{Knill2001} and entanglement measurements \cite{Bouwmeester1997} are implemented. 
Group-velocity dispersion can lead to temporal broadening and distortion of a single-photon pulse shape \cite{Valencia2002, Brida2006, Baek2008, Li2019, Jin2024}, reducing the indistinguishability of individual single photons and their two-photon interference visibility \cite{Fan2021}. 
Therefore, preparing pure single photons in the low-dispersion telecom O-band can not only extend the available wavelength range but also contribute to the realization of high-precision photonic quantum gate operations in quantum computing and quantum network applications. 
For example, pure single photons at the low-dispersion wavelength will be beneficial in improving the bandwidth of the measurement-device-independent quantum key distribution (MDI-QKD) \cite{Lo2012, Lo2014}, where secure keys are generated via two-photon interference in a middle-point station with security against detector-side-channel attacks. 
Pure and indistinguishable single photons can be transformed into entangled photons by using two-photon interference and linear optical quantum gate operations \cite{Knill2001}. 
Such entangled photons at the zero-dispersion dispersion wavelength can be a resource for quantum repeater networks, which also require two-photon interference in entanglement swapping. 
Although the low-dispersion propagation can be achieved by dispersion-shifted fibers and the combination of fibers having positive and negative dispersions at other communication bands, standard optical fibers are predominantly used in practical applications. Thus, single-photon sources at the O-band can be deployed without system modification and benefit from the low-dispersion feature.

Furthermore, broadening the communication bands of pure single photons enables the use of the nonlinearity of optical fiber cables. 
By employing largely detuned wavelengths, one can utilize the desired nonlinear processes while minimizing Raman scattering, a major source of noise photons in fiber cables. 
For example, a Bragg-scattering four-wave-mixing process pumped by C-band laser sources has been demonstrated for switching the wavelength of single photons at the O-band \cite{Joshi2018}.
The multi-mode nonlinear switching scheme has also been applied to the frequency multiplexing of heralded single-photon states at the O-band. 
An all-optical ultrafast single-photon switching has also been demonstrated using the O-band \cite{Hall2011}.
All-optical storage has been achieved by nonlinear wavelength conversion in a fiber cavity quantum memory \cite{Bonsma2024}. 
Incorporated with pure heralded single photons at the O-band, those advanced all-optical techniques will be truly valuable technologies for quantum state synthesis and quantum network applications. 

Recent advancements have led to the development of several heralded single-photon sources at the telecom O-band using spontaneous parametric downconversion (SPDC) in periodically poled lithium niobate waveguides \cite{Martin2009, Bock2016} and periodically poled silica fibers \cite{Chen2021}. 
Single-photon generation at the O-band has also been realized by semiconductor quantum dots \cite{Kolatschek2021}.
High indistinguishability (96\%) of sequentially produced single photons from one quantum dot source has been achieved in \cite{Srocka2020}. 
However, those experiments have not demonstrated the generation of indistinguishable photons from independent sources. 
Pure heralded single photons can be prepared by tight spectral filtering; however, this significantly diminishes the source brightness and the heralding efficiency, i.e., the generation and collection probability of a single photon conditional on the detection of its twin photon \cite{Evan2017}.
Therefore, the direct generation of pure heralded single photons is essential.
%\textcolor{red}{
%We observed that previous studies have generally lacked high-quality photon sources in the O-band; inspired by these findings, we have undertaken research on the direct generation of pure heralded single photons in the O-band.}

In this paper, we demonstrate the numerical optimization of spectrally pure heralded single-photon sources at the telecom O-band, utilizing group-velocity-matching (GVM) conditions and custom poling structures of a nonlinear optical crystal. 
Engineered GVM conditions \cite{Grice2001} have been widely applied for heralded single-photon sources at the telecom C-band \cite{Evans2010, Harder2013, Bruno2014}, L-band \cite{Jin2013, Weston2016, Kaneda2016}, and 800-nm range \cite{Mosley2008PRL}, but not yet at the telecom O-band. 
We investigate a range of possible SPDC wavelength configurations in which a heralded single photon is produced at 1310 nm with suitable GVM conditions in a potassium titanyl phosphate (KTP) crystal that has been successfully used for high-purity single photons at the telecom C-band and L-band. 
A poling structure and its associated second-order optical nonlinearity of a KTP crystal is then designed 
to shape a phase-matching function (PMF) to a near-Gaussian form. 
A conventional periodically-poled (PP) structure produces a sinc-shaped PMF with undesired spectral lobes degrading spectral purity. 
In this paper, we employ recently demonstrated poling optimization schemes with the unit poling lengths of the phase-matching coherence length (CL) and sub-coherence length (SCL) \cite{Tambasco2016, Dosseva2016, Graffitti2017, Graffitti2018optica, Pickston2021, Zhang2022, Baghdasaryan2023}. In the optimization procedure, we also take into account poling domain length, pump wavelength and bandwidth, and wavelengths of photon pairs to find sources feasible and reproducible with current and near-future technologies. 
We also investigate undesired noise spectra of our designed sources and their effect on spectral purity. 

%%%%%%%%%%%%%%%%%%%%%%%%%%%%%%%%%%%%%%%%%%%%%%%%%%%%

\section{Method}
\subsection{Joint spectral amplitude of SPDC photons}

\begin{figure}[tbp]
\centering
\includegraphics[width=1 \columnwidth]{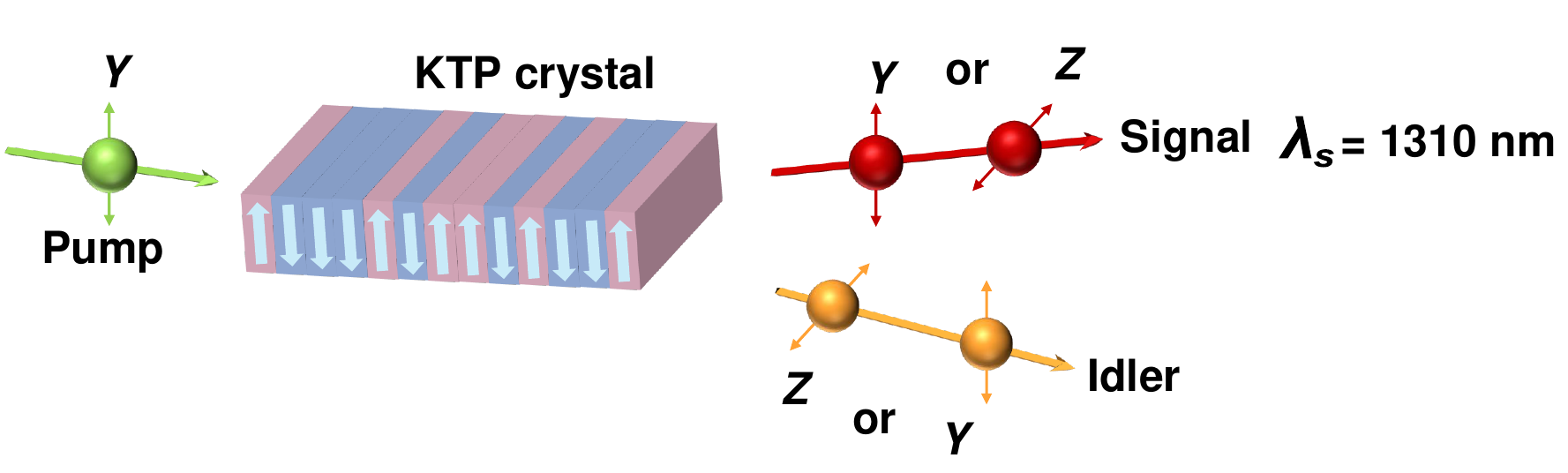}
\caption{Conceptual diagram of a spectrally pure heralded single-photon source at the telecom O-band optimized by an appropriate GVM condition and a modulated poling structure of a potassium titanyl phosphate (KTP) crystal. We consider a Type-II quasi-phase-matching condition where pump, signal, and idler photons are polarized along the crystallographic $Y$, $Z(Y)$, and $Y(Z)$ axes, respectively. }
\label{concept}
\end{figure}

A conceptual diagram of a heralded single-photon source to be optimized in this paper is depicted in Fig. \ref{concept}. As a nonlinear medium, we consider a KTP crystal satisfying a Type-II collinear quasi-phase-matching condition. A pump pulse is polarized along the crystallographic $Y$ axis, generating two SPDC photons polarized parallel to the $Y$ and $Z$ axes. In this paper, we will refer to one SPDC photon, which should have a wavelength of 1310 nm, as the signal photon and the other as the idler photon. When a signal photon has $Y$($Z$) polarization, its twin idler photon has $Z$($Y$) polarization. Angular frequency and wavenumber are represented by $\omega_j$ and $k_j$, where $j$ denotes pump ($j = p$), signal ($j = s$), and idler ($j=i$) modes, respectively. Under the plane-wave approximation for the pump, signal, and idler modes, the two-photon joint spectral state $\vert\psi\rangle$ without a normalization factor is given by
\begin{equation}\label{SPDC}
\vert\psi\rangle=\int_0^\infty\int_0^\infty\,\mathrm{d}\omega_s\,\mathrm{d}\omega_if(\omega_s,\omega_i)\hat{a}_s^\dag(\omega_s)\hat{a}_i^\dag(\omega_i)\vert0\rangle_s\vert0\rangle_i,
\end{equation}
where $\hat{a}_j^\dag (\omega_j)$ is the creation operator of the mode $j$. The joint spectral amplitude (JSA) of an SPDC two-photon state $f(\omega_s,\omega_i)$ is characterized by the pump-envelope function $\alpha(\omega_s,\omega_i)$ and the PMF $\phi(\omega_s,\omega_i)$: $f(\omega_s,\omega_i) = \alpha(\omega_s,\omega_i)  \phi(\omega_s,\omega_i)$.
%A joint spectral amplitude (JSA) of an SPDC two-photon state $f(\omega_s,\omega_i) $ is calculated by the product of a pump-envelope function $\alpha(\omega_s,\omega_i)$ and PMF $\phi(\omega_s,\omega_i)$, i.e., $f(\omega_s,\omega_i) = \alpha(\omega_s,\omega_i)  \phi(\omega_s,\omega_i)$.
We assume a pump envelope function with a Gaussian distribution,
\begin{equation}\label{PEF}
\alpha(\omega_s, \omega_i)=\exp\left[-\left(\frac{\omega_s+\omega_i-\omega_{p_0}}{\sigma_p}\right)^2\right],
\end{equation}
where $\omega_{p_0}$ and $\sigma_p$ denote a central frequency and a bandwidth of a pump pulse. 
The anti-correlation of signal and idler frequencies in $\alpha(\omega_s, \omega_i)$ represents the energy conservation in the SPDC process.  
A PMF for collinear, plane-wave SPDC is given by
\begin{equation}\label{PMF_general}
\begin{aligned}
\phi \left( \Delta k\left( \omega _s,\omega _i \right), z \right) = \int_0^z{dz' g\left( z' \right)}e^{i \Delta k\left( \omega _s,\omega _i \right) z'}.
\end{aligned}
\end{equation}
Here, $z$ is the position in the crystal having the length $L$ ($0 \leq z \leq L$). We utilize $\phi(\Delta k, z)$ obtained by integrating a part of the crystal ($z < L$) in the optimization process of $\phi(\Delta k, L)$ produced by the entire crystal.
The phase mismatch $\Delta k = k_p-k_s-k_i$ depends on the material dispersion with $k_j=\omega_j n_j(\omega_j)/c$, where $n_j(\omega_j)$ is the refractive index as a function of the angular frequency and $c$ is the speed of light in vacuum. 
The normalized nonlinearity function $g(z)$ can be $+1 (-1)$ for the poling orientation UP (DOWN). 
For example, a bulk nonlinear crystal has $g(z)=1$ for $0 \leq z \leq  L$, and $\Delta k$ needs to be zero to achieve the maximum generation efficiency of SPDC. 
A PP scheme satisfying a 1st-order quasi-phase-matching condition can compensate for the phase mismatch by the crystal's poling orientations alternating between UP and DOWN with a period of phase-matching coherence length $l_c = \pi/ |\Delta k|$. 
The PMF produced by the PP scheme is given by
\begin{equation}\label{PMF_PP}
\begin{aligned}
\phi_{\textrm{PP}}(\Delta k(\omega_s, \omega_i))&= 
\frac{2}{\pi}\mathrm{sinc}\left[ \left( \Delta k(\omega_s, \omega_i) - \frac{\pi}{l_c} \right)\frac{L}{2}\right]\\
&e^{i\Delta k(\omega _s,\omega _i)\frac{L}{2}}.
\end{aligned}
\end{equation}

%\begin{equation}\label{PMF_PP}
%\begin{aligned}
%\phi_{\textrm{PP}}(\Delta k(\omega_s, \omega_i))
%=\frac{2}{\pi}\mathrm{sinc}\left[ \left( \Delta k(\omega_s, %\omega_i) - \frac{\pi}{l_c} \right)\frac{L}{2}\right]e^{i\Delta %k(\omega_s, \omega_i)\frac{L}{2}}.
%\end{aligned}
%\end{equation}
%
%The PMF has the maximum generation efficiency for $\Delta k = \pi / l_c$. 
The sinc distribution in Eq. (\ref{PMF_PP}) has peripheral lobes that degrade the purity of heralded single photons (a sinc-shaped PMF is also produced by a bulk crystal).
Our goal is to shape the PMF to an optimal Gaussian form \cite{Quesada2018, Graffitti2018} for different GVM cases by modulating a poling structure of a KTP crystal. 

Spectral purity of heralded a single-photon state $P$ can be calculated by using Schmidt decomposition on  JSA \cite{Mosley2008},
\begin{equation}\label{schmidt}
f(\omega_s,\omega_i) = \sum_jc_j\phi_j(\omega_s)\varphi_j(\omega_i),
\end{equation}
\begin{equation}\label{purity}
P=\sum_jc_j^4,
\end{equation}
where $\phi_j(\omega_s)$ and $\varphi_j(\omega_i)$ are orthogonal spectral amplitude distributions of the signal and idler modes, and $c_j$ is a set of non-negative real numbers that satisfy the normalization condition $ \sum_jc_j^2=1$.
In our numerical calculation, since JSA is obtained in a discretized matrix form, the Schmidt decomposition is performed by singular value decomposition.

\subsection{Group-velocity matching}
As shown in Eq. (\ref{PMF_PP}), the phase mismatch is a crucial parameter in determining the PMF. 
For a Type-II SPDC source (which is of our interest), $\Delta k(\omega_s, \omega_i)$ is well approximated by the first-order dispersion: $\Delta k(\omega_s, \omega_i) \sim \Delta k_0 + k'_p (\Omega_s + \Omega_i) -k'_s \Omega_s -k'_i \Omega_i$, where $\Delta k_0$ is the phase mismatch at the center frequencies of pump, signal, and idler modes. $\Omega_j = \omega_j - \omega_{j0}$ is the frequency detuning from the central frequency $\omega_{j0}$ and $k'_j = \frac{dk_j}{d\omega_j}\big|_{\omega_{j0}}$ is the inverse group velocity at $\omega_{j0}$. 
Since the PMF has the maximum value for $\Delta k(\omega_s, \omega_i) = \Delta k_0$ with a quasi-phase-matching condition, an orientation of $\phi (\Delta k)$ can be characterized by the GVM angle:

\begin{equation}\label{D}
\theta = \arctan \qty(\frac{\Omega_i}{\Omega_s} )= \arctan \qty(-\frac{k'_p-k'_s}{k'_p-k'_i} ).
\end{equation}
Since the signal and idler frequencies are anti-correlated in the pump envelope function, the correlation in the PMF should be positive, i.e., $0^{\circ} \leq \theta \leq 90^{\circ}$, to achieve separable JSA and thereby a pure heralded single-photon state. 
%The representative GVM angles of $45^{\circ}$ \cite{Grice2001,konig2004} and $0^{\circ}(90^{\circ})$ \cite{Kaneda2016} have been demonstrated at the telecom C-band and L-band. 
We perform a numerical search for positively correlated PMFs producing signal photons at 1310 nm.

\subsection{Poling optimization}
\begin{figure}[t!]
\centering
\includegraphics[width=0.48\textwidth]{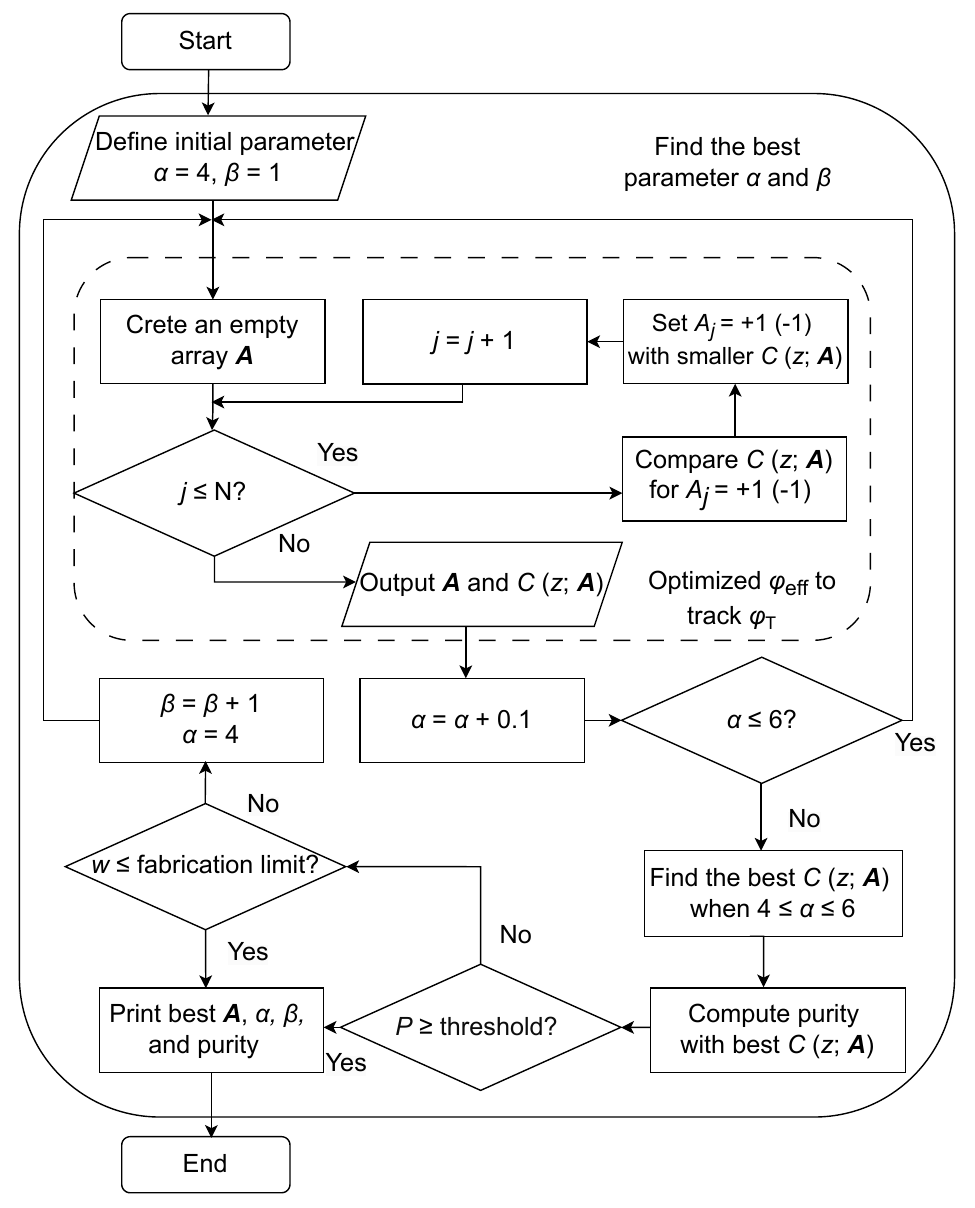}
\caption{
Flowchart of the poling optimization algorithm. The poling array $\vb*A$ is constructed for given parameters $\alpha$ and $\beta$, and then evaluated by the cost function $C(z; \vb*{A})$, where $\alpha$ is the Gaussian spatial width factor and $\beta$ is the domain division factor. The procedure is continued until the poling array shows heralded single-photon purity higher than the threshold ($P = 99.5$\%).
}
\label{CPKTP-algorithm}
\end{figure}

In order to achieve the generation of separable JSA and a pure heralded single photon, one needs appropriate GVM conditions and a Gaussian PMF as demonstrated in \cite{Quesada2018, Graffitti2018}. 
The ideal Gaussian PMF is given by 
\begin{equation}\label{phi_I_k_main}
\phi_{\textrm{I}}(\Delta k) = \exp \left[-\frac{1}{2}\left(\Delta k-\Delta k_0\right)^{2}\sigma^{2}\right],
\end{equation}
where $\sigma$ is the spatial width of a Gaussian nonlinearity profile.
Due to the Fourier relationship as shown in Eq. (\ref{PMF_general}), the ideal spatial nonlinearity $g_{\textrm{I}} (z)$ producing $\phi_{\textrm{I}} (\Delta k)$ is also in a Gaussian form. 
However, such $g_{\textrm{I}} (z)$ cannot be achieved by a practical nonlinear crystal due to the limited crystal length and the binary adjustment of the nonlinearity. 
To produce the PMF closely approximated to the Gaussian form, we employ the combination of the CL and SCL poling modulation schemes, each of which has been independently demonstrated in previous works \cite{Graffitti2017, Graffitti2018optica, Zhang2022}. The CL and SCL schemes respectively produce a poling structure with the unit poling domain length $w= l_c$ and $ l_c/\beta$, where $\beta >1$ is a domain division factor. Thus, compared with the CL scheme, the SCL scheme has a larger number of unit poling domains $N = \floor{L/w} = \floor{L\beta/l_c}$ for the more precise adjustment of the spatial nonlinearity, where $\floor{x}$ gives the greatest integer less than or equal to x. The poling optimization procedure is described as follows: 
We first set the target PMF $\phi_{ \textrm{T}}(\Delta k = \Delta k_0, z)$, which is obtained by the integral of Eq. (\ref{PMF_general}) from $z' = 0$ to $z$ $(\leq L)$ with $g(z') = g_{\textrm{I}}(z')$ and $\Delta k = \Delta k_0 = \pi/l_c$ (see Appendix A for the details of the derivation): 
\begin{equation}\label{phi_T_z_main}
\begin{aligned}
\phi_{ \textrm{T}}(\Delta k = & \Delta k_0,z) = 
\\& \sqrt{\frac{2}{\pi}}\sigma\left[\operatorname{erf}\left(\frac{L}{2\sqrt{2} \sigma}\right)+\operatorname{erf}\left(\frac{z-L/2}{\sqrt{2} \sigma}\right)\right],
\end{aligned}
\end{equation}
where $\operatorname{erf}()$ denotes the error function. 
The target PMF is defined for an arbitrary position $z$ in the crystal so that $\phi_{ \textrm{T}}(\Delta k = \Delta k_0,z)$ can be compared with the designed PMF created by a part of poling region, where $g(z)$ can be $\pm1$. At $z = L$, $\phi_{ \textrm{T}}(\Delta k = \Delta k_0,z)$ is the closest to $\phi_{ \textrm{I}}( \Delta k = \Delta k_0)$. 
The designed effective PMF produced by the 1-st to $n$-th poling domains ($n = $ $\floor{z/w}$ $\leq N$) is given by 
\begin{equation}\label{phi_eff_z_main}
\begin{aligned}
\phi_{\textrm{eff}}( \Delta k = \Delta k_0,  z & = nw; \vb*{A}) =\sum_{j=1}^{n} A_j \int_{(j-1)w}^{jw} e^{i\Delta kz'}dz'
\\ & =\frac{i}{\Delta k_0}(e^{-iw \Delta k_0}-1)\sum_{j=1}^{n}{A_je^{iwj\Delta k_0}},
\end{aligned}
\end{equation}
where $\vb*{A}$ and $A_j$ are the domain orientation array and its $j$-th element $(A_j=+1,-1)$ representing the orientation of the $j$-th domain. 
The closeness of $\phi_{\textrm{eff}}$ to $\phi_{\textrm{T}}$ is evaluated by the cost function $C(z; \vb*{A})$: \begin{equation}\label{cost_main}
\begin{aligned}
C(z; \vb*{A})=\abs{{\phi_{\textrm{eff}}(z;\vb*{A})-\phi_\textrm{{T}}(z)}}^{2}.
\end{aligned}
\end{equation}

In our optimization procedure, $\vb*{A}$ is optimized together with the Gaussian spatial width $\alpha= L/\sigma$ and poling domain division
factor $\beta=  l_c/w$, which respectively determine $\phi_{\textrm{T}}(z)$ and the optimization precision of $\phi_{\textrm{eff}}(z)$. 
A larger $\alpha$ can result in a more precise approximation with a Gaussian function at the cost of lower brightness. 
In previous works, this trade-off value has been chosen as $\alpha = 4.7$  \cite{Pickston2021} and $\alpha = 5$ \cite{Graffitti2021thesis}. 
In order to explore the possibility of further optimization of single-photon purity, we compute the cost function to find the best value of $\alpha$ for achieving the highest purity within the range of $4 \leqslant \alpha \leqslant 6$ with a step size of 0.1. 
As discussed above, $\beta$ is a parameter to adjust the number of poling domains. In practice, however, $w$ cannot be arbitrarily small due to the limited precision of poling techniques by the presence of domain walls and the uncertainty in their placement. Thus, in this work, we search for a poling structure achieving $P > 99.5$\% with the smallest $\beta$. 

The flowchart of the algorithm to optimize $\alpha$, $\beta$, and $\vb*{A}$ is illustrated in Fig. \ref{CPKTP-algorithm}. 
First, for given $\alpha$ and $\beta$ (which are initially set at 4 and 1), elements of $\vb*{A}$ are sequentially determined from $j = 1$ to $N$ so that $C(z=jw; \vb*{A})$ is minimized;  
$A_j$ is chosen to be $+1(-1)$ when $C(z; \vb*{A})$ for $A_j=+1$ is less (larger) than that for $A_j=-1$. 
The procedure is repeated to optimize the poling arrays for $4 \leqslant \alpha \leqslant 6$ (and a fixed value of $\beta$), and the one having the lowest cost function is selected. 
Heralded single-photon purity $P$ is then calculated for JSA, which is obtained as a product of $\phi_{\textrm{eff}}$ produced by the selected array and a pump envelope function $\alpha (\omega_s, \omega_i)$ with an optimal pump bandwidth. 
If the purity does not meet the threshold (99.5\%), the optimization loop is repeated after changing $\beta$. 
The optimization procedure continues until the purity exceeds the threshold or $w = l_c/\beta$ reaches less than 1 $\mu$m \cite{Graffitti2021thesis}, which is about the limit with the current fabrication technology.

%%%%%%%%%%%%%%%%%%%%%%%%%%%%%%%%%%%%%%%%%%%%%%%%%%%%

\section{Optimization of GVM and crystal poling structure}

\begin{figure*}[t!]
\centering\includegraphics[width=0.95\textwidth]{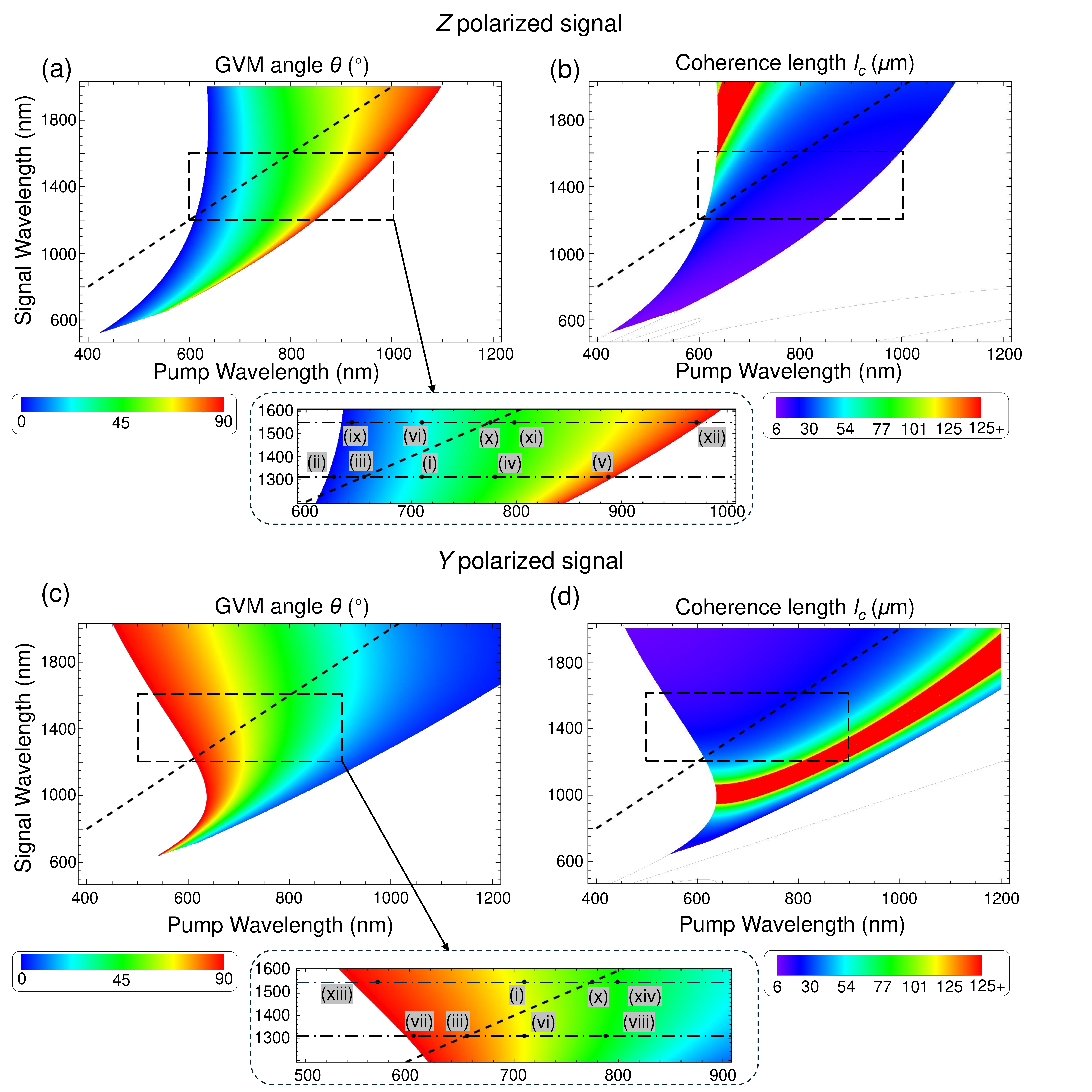}
\caption{(a,c) GVM angle $\theta$ and (b,d) phase-matching coherence length $l_c$ as a function of pump and signal wavelengths in a KTP crystal. The signal mode is (a,b) $Z$-polarization and (c,d) $Y$-polarization. 
The GVM angle distribution is mapped for $0^{\circ} \leq \theta \leq 90^{\circ}$, where high heralded single-photon purity can be achieved. Dashed black lines indicate the degenerate case $\lambda_s = 2\lambda_p$. The insets of (a,c) show our selected GVM cases to perform poling optimization at $\lambda_s = $ 1310 nm and $\lambda_s = $ 1550 nm (see Appendix C for $\lambda_s = $ 1550 nm).  
}
\label{GVM-theta}
\end{figure*}

To explore possible configurations for generating pure heralded single photons at the telecom O-band by using a KTP crystal, we calculated the GVM angle $\theta$ as a function of the pump wavelength $\lambda_p$ and the signal wavelength $\lambda_s$ \cite{Mccracken2018}.  Figure \ref{GVM-theta}(a) and (c) show the GVM angle $\theta$ for the $Z$- and $Y$-polarized signal photons, respectively. The GVM angle is plotted for $0^{\circ} \leq \theta \leq 90^{\circ}$, the range required for high heralded single-photon purity. The dashed black line represents degenerate SPDC ($\lambda_s = \lambda_i$). The phase-matching coherence length $l_c$ for the $Z$- and $Y$-polarized signal photons is shown in Fig. \ref{GVM-theta} (b) and (d), respectively. For $0^{\circ} \leq \theta \leq 90^{\circ}$, $l_c$ is over $11 \mu$m, which is available with current poling technology.

We select several GVM cases to perform poling optimization. 
As shown in the inset of Fig. \ref{GVM-theta}(a), we take five cases producing a $Z$-polarized signal photon: (\romannumeral1) $\theta = 26^{\circ}$ and $\lambda_i = 1550$ nm (telecom C-band); (\romannumeral2) $\theta = 2^{\circ}$; (\romannumeral3) $\theta = 10^{\circ}$ and $\lambda_i =1310$ nm (degenerate case); (\romannumeral4) $\theta = 45^{\circ}$; (\romannumeral5) $\theta = 85^{\circ}$. 
We also consider three cases of $Y$-polarized signal photons, as shown in Fig. \ref{GVM-theta} (c): (\romannumeral6) $\theta = 67^{\circ}$ and $\lambda_i = 1550$ nm; (\romannumeral7) $\theta = 89^{\circ}$ and (\romannumeral8) $\theta = 45^{\circ}$. 
Note that the GVM angle $\theta \sim 0^{\circ}$ is not available for the $Y$-polarized signal photon since the corresponding idler photon has $\lambda_i > 4$ $\mu$m, outside the transparency range of a KTP crystal \cite{Kato1991}. 
In the GVM cases (\romannumeral2,\romannumeral5,\romannumeral7), we chose $\theta$ slightly deviated from 0$^\circ$ and 90$^\circ$ so that the spectral purity can be maximized by using a finite pump bandwidth \cite{Kaneda2016}. Note that, in the Fig. \ref{GVM-theta} (a) and (b), the repeated Roman numerals refer to the same GVM case.
\begin{figure*}[t!]
\centering\includegraphics[width=0.9\textwidth]{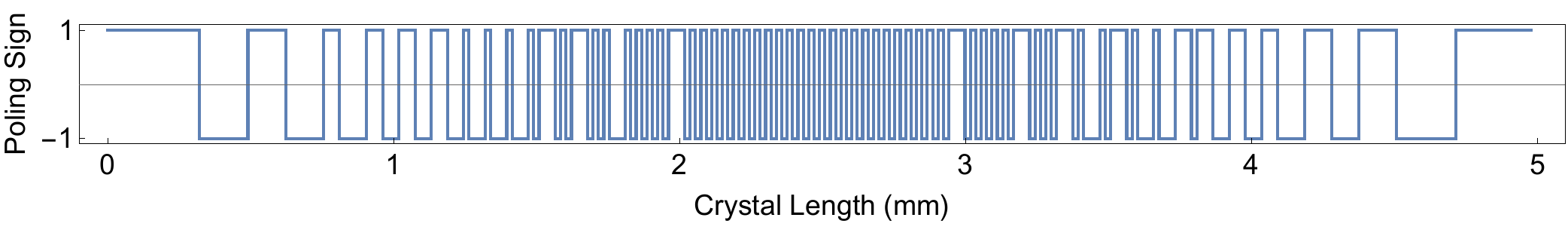}
\caption{Optimized poling structure of a KTP crystal for the GVM case (i) ($\theta = 26^{\circ}$, $\lambda_i = 1550$ nm). The crystal poling signs $+1$ and $-1$ denote UP and DOWN poling directions, respectively.}
\label{710-purity}
\end{figure*}

\begin{table*}[t!]
\centering
\caption{Summary of the optimized parameters and purity for the selected GVM cases (i-viii) producing heralded single photons at $\lambda_s = $ 1310 nm. The signal photon is $Z$-polarized for the cases (\romannumeral1-\romannumeral5) and $Y$-polarized for the cases (\romannumeral6-\romannumeral8). 
$P_{P}$ and $P_O$ show the purity for the PP and (S)CL poling schemes, respectively. The optimized pump bandwidth $\Delta\lambda_p$ is used when $P_O$ is calculated. 
}
%\begin{tabularx}{\textwidth}{ccccccccc}
\begin{tabular}{c c c c c c c c c}
\hline \hline
Case & $\lambda_y \rightarrow \lambda_z+\lambda_y$ & $\theta$ & $l_c$ ($\mu$m) & $\alpha$ & $\beta$ & $\Delta\lambda_p$ (nm) &$P_P$& $P_O$ \\
\hline
(\romannumeral1)&$ 710.0 \rightarrow 1310 + 1550 $ & 26$^{\circ}$ & 18.86 & 5.1 & 1& 3.07 & 83.01\% & 99.49\% \\
(\romannumeral2)&$ 626.3 \rightarrow 1310 + 1200 $ & 2$^{\circ}$ & 39.50 & 6 & 18 & 8.13 & 93.20\% & 99.57\% \\
(\romannumeral3)&$ 655.0 \rightarrow 1310 + 1310 $ & 10$^{\circ}$ & 27.33 & 4.3 & 4 & 3.03 & 86.16\% & 99.55\% \\
(\romannumeral4)&$ 779.5 \rightarrow 1310 + 1925 $ & 45$^{\circ}$ & 14.99 & 5.4 & 1 & 3.74 & 82.32\% & 99.52\% \\
(\romannumeral5)&$ 887.3 \rightarrow 1310 + 2750$ & 85$^{\circ}$ & 13.18 & 5.6 & 10 & 11.9 & 88.92\% & 99.41\% \\
\hline
(\romannumeral6)&$ 710.0 \rightarrow 1550 + 1310 $ & 67$^{\circ}$ & 31.40 & 5.3 & 4 & 3.62 & 83.30\% & 99.52\% \\
(\romannumeral7)&$ 603.8 \rightarrow 1120 + 1310 $ & 89$^{\circ}$ & 25.84 & 5.9& 18 & 7.80 & 94.49\% & 99.76\% \\
(\romannumeral8)&$ 787.8 \rightarrow 1977 + 1310 $ & 45$^{\circ}$ & 44.71 & 5.3 & 6 & 4.16 & 82.31\% & 99.68\% \\
\hline \hline
\end{tabular}\\
\label{result-1310}

\end{table*}

\begin{figure*}[tbp]
\centering\includegraphics[width=0.9\textwidth]{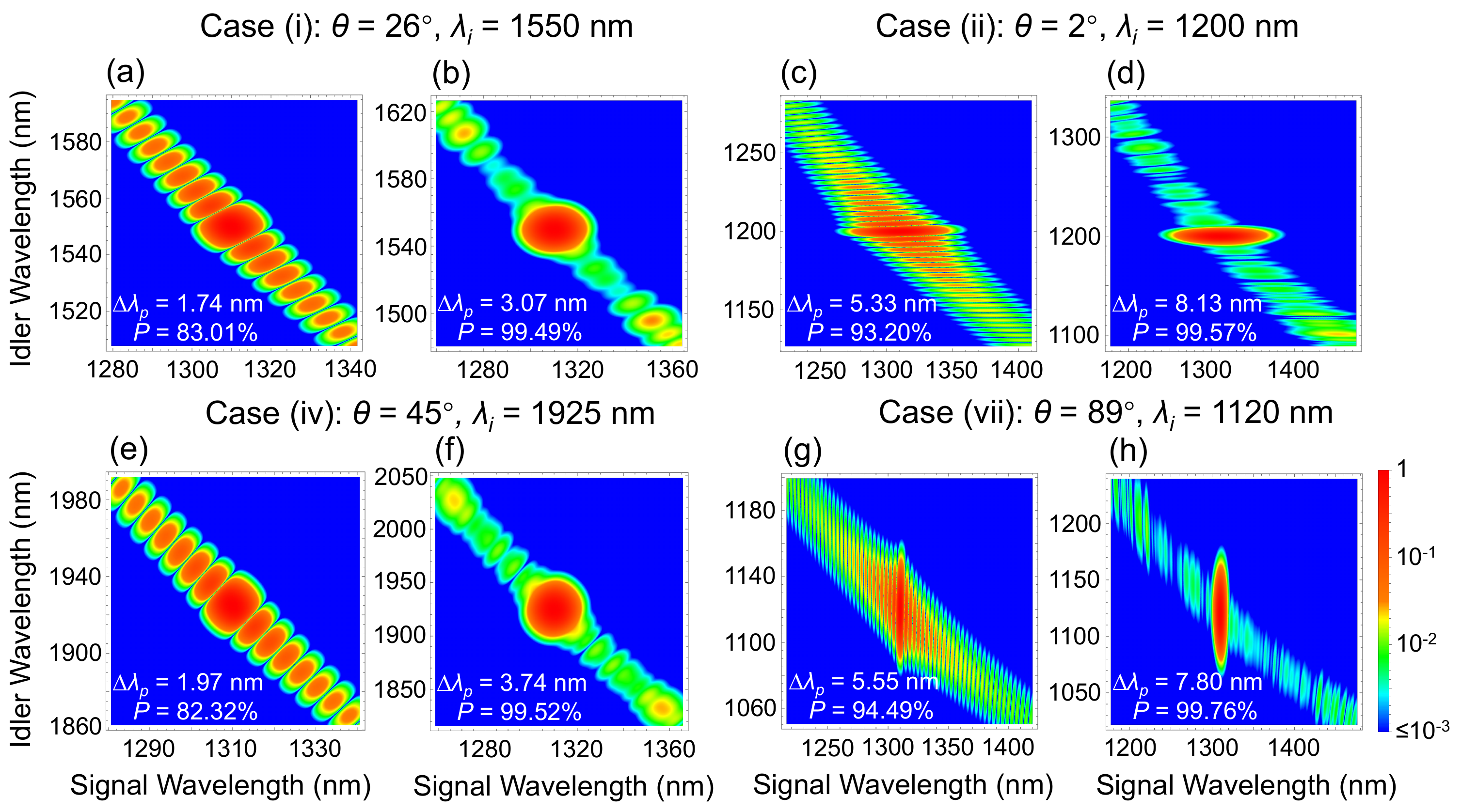}
\caption{JSA for the GVM cases (\romannumeral1,\romannumeral2,\romannumeral4,\romannumeral7). (a,c,e,g) PP scheme. (b,d,f,h) CL and SCL schemes. $\lambda_i$, the center wavelength of the idler mode; $\Delta\lambda_p$, optimal pump bandwidth achieved the maximum purity with the corresponding PMF. 
}
\label{GVM}
\end{figure*}

\begin{figure*}[t!]
\centering\includegraphics[width=0.9\textwidth]{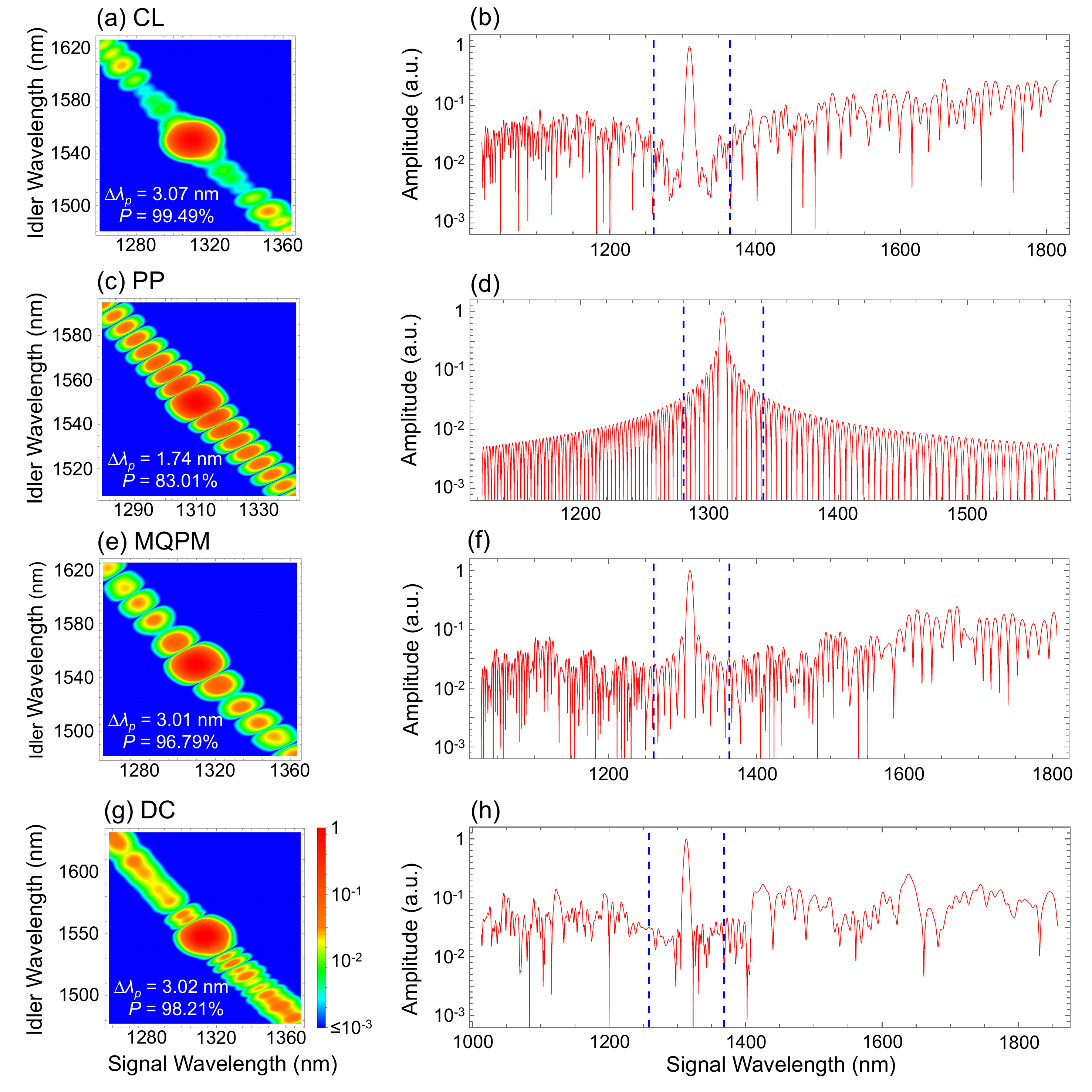}
\caption{JSA for the GVM case (i) mapped in the spectral range of $R = 10 \Delta\omega$ (left) and its frequency anti-correlation distribution $| f ( \omega_s, \omega_{s0} + \omega_{i0} -\omega_s )|$ in the spectral range of $R = 70 \Delta \omega$ (right). For comparison, we show plots for the different poling schemes: (a,b) CL scheme (which we employed in this work), (c,d) PP scheme, (e,f) multi-order quasi-phase-matching condition (MQPM) modulation scheme, and (g,h) duty-cycle (DC) modulation scheme. 
The area between the two dashed lines in (b,d,f,h) show the range of $R = 10 \Delta\omega$.
$\Delta \lambda_p$, optimal pump bandwidth; $P$, spectral purity.} 
\label{wide range}
\end{figure*}

For each selected GVM case, an optimal poling structure and the corresponding PMF are searched for the crystal length of $L = 5$ mm. 
An optimized poling structure for case (\romannumeral1) is shown in Fig. \ref{710-purity} as an example. 
Our optimization result for the GVM cases (\romannumeral1-\romannumeral8) is summarized in Table \ref{result-1310}. 
Optimized JSA for representative GVM cases (\romannumeral1,\romannumeral2,\romannumeral4,\romannumeral7) are shown in Fig. \ref{GVM}.
For all GVM cases, the peripheral lobes are highly suppressed by the CL and SCL poling structures. With the optimal pump bandwidth $\Delta \lambda_p$, the spectral purity for the optimized poling structures $P_O$ is over 99\%, superior to that of PP structures $P_P$ ($82-95\%$).
Each optimized JSA exhibits a wider phase-matching bandwidth and pump bandwidth $\Delta \lambda_p$ compared to the corresponding JSA produced by the PP structure. This is attributed to the shorter effective crystal length in the optimized poling structures that reduce the region of high nonlinearity.  
We note that by applying similar optimization procedures, high-purity sources have also been designed at 1550 nm (see Appendix C).

In the calculation of the purity from the JSA, which is produced in a matrix form, we carefully chose the spectral range $R$ and resolution $D$ \cite{Graffitti2018, Graffitti2021thesis}.
If the spectral range is too small, unwanted peripheral lobes in JSA can be filtered out, and therefore, one can overestimate the purity. 
A spectral resolution should also be sufficient to correctly calculate the contribution from unwanted, narrowband peaks (at the cost of a longer computational time). 
For the cases (\romannumeral1,\romannumeral3,\romannumeral4,\romannumeral6,\romannumeral8) ($10^\circ \leq \theta \leq 80^\circ$) in which the signal bandwidth is comparable to the idler one (less than 2.25 times), we calculated the purity with of $R = 10\Delta \omega$ and $D = \Delta \omega /20$ (and thus JSA is generated as a $200 \times 200$ matrix). 
Here $\Delta \omega$ denotes the average bandwidth (full width at half maximum) of the signal and idler modes at the central peak of JSA.
The ratio $R/\Delta \omega$ is an optimized value for a sufficiently precise estimation of the purity in the previous work \cite{Graffitti2018}, and the number of matrix elements in our calculation is four times larger than that in \cite{Graffitti2018}. 
For the cases (\romannumeral2,\romannumeral5,\romannumeral7) ($\theta < 10^\circ$ or $\theta > 80^\circ$),
the bandwidth of one SPDC mode is much narrower than the other, and peripheral lobes also have narrow bandwidth, as observed in Fig. \ref{GVM}(d) and (h). Therefore, we chose $R = 10 \Delta \omega$ and $D = \Delta \omega /40$ (each JSA is generated as a $400 \times 400$ matrix) for the fine discretization of JSA and precise estimation of the purity.  
As will be discussed, $R$ can also be a simulation of a photon collection bandwidth, which is a practically important parameter for achieving high spectral purity.

We see that the optimized domain division factor $\beta$ is highly dependent on $\theta$ and $l_c$; for the cases (\romannumeral2,\romannumeral5,\romannumeral7) ($\theta \approx 0^{\circ}, 90^{\circ}$), the peak bandwidths of the signal and idler modes are highly asymmetric, and many lobes are placed within the wider of the signal or idler peak bandwidth, as shown in Fig. \ref{GVM} (c,e). In such cases, the spatial nonlinearity needs to be precisely controlled with $\beta > 10$ to eliminate the lobes. 
The SCL scheme ($\beta > 1$) also helps to improve the optimization precision when $l_c$ is too large to achieve $> 99$\% purity with the CL scheme ($\beta = 1$). The GVM cases (\romannumeral3,\romannumeral6,\romannumeral8) with $l_c > 25$ $\mu$m are optimized with the SCL structures ($\beta > 1$), while the cases (\romannumeral1, \romannumeral4) with $10^{\circ} \leq \theta \leq 80^{\circ}$ and $l_c < 20$ $\mu$m, are optimized with $\beta = 1$. Note that we have attempted cases for $\theta = 0^{\circ}, 90^{\circ}$. However, the poling structures for those cases are optimized with $\beta > 50$, and $w < 1$ $\mu$m, which is not practical with current fabrication capabilities. 
The ratio of the crystal length to the target Gaussian width $\alpha$ is not clearly correlated with the other parameters. 
The deviation of $\alpha$ may be due to the algorithm's limited tracking precision. 
Since there is a finite number of domains, each target PMF may accidentally align with a specific value of $\alpha$, minimizing the cost function and thus maximizing purity. 
Indeed, we see that the choice of $\alpha$ has a smaller impact on the purity as shorter domain width $w$.
For instance, in the GVM case (ii) with $\beta = 10$, the difference between the maximum and minimum purity values for $4 \leqslant \alpha \leqslant 6$ is only about $0.3\%$, whereas the difference increases to $1.7\%$ for $\beta = 4$ and $7.1\%$ for $\beta = 1$.
In this sense, parametrizing $\alpha$ helps to achieve high purity with limited poling precision.

Our designed sources are feasible with current off-the-shelf and state-of-the-art technologies. 
In particular, in the GVM cases (\romannumeral1) and (\romannumeral6), pure heralded single photons at 1310 nm can be produced by the detection of its twin photon in the telecom C-band (1550 nm), where commercial superconducting nanowire single-photon detectors (SNSPD) can be operated with a $>$ 90\% detection efficiency. In the telecom O-band and C-band, a detection efficiency of $> 98$\% has been achieved experimentally \cite{Reddy2020, Chang2021}. 
A mode-locked Ti:sapphire laser can be an appropriate pump source for the two cases. 
Note that the pump wavelengths of our designed sources range from 603.8 nm to 887.3 nm, which can be fully supported by off-the-shelf laser sources such as Ti:sapphire lasers and frequency-doubled ultrafast optical parametric oscillators pumped by the Ti:sapphire lasers and mode-locked Yb-doped fiber lasers. 
The cases (\romannumeral4,\romannumeral5,\romannumeral8) require the detection of single photons in the mid-infrared band, which can be achieved by advanced SNSPDs with extended detection wavelength range \cite{Dello2022, Prabhakar2020}. 

The poling fabrication is also an important factor in feasibility. The cases (i,iii,iv,vi,viii) require the minimum poling length of $> 5$ $\mu$m. Given that KTP crystals with down to $\sim 3$  $\mu$m poling periods have been successfully fabricated and applied to various wavelength conversion modules and SPDC sources, the deviation of poling domain lengths is estimated to be much less than 1\%. Thus, there are no technical obstacles to fabricating poling structures for those GVM cases. In \cite{Graffitti2018}, the degradation of purity for $< 1$\% poling length deviations is to be negligibly small for the CL scheme at the C-band. 
The cases (\romannumeral2,\romannumeral5,\romannumeral7) have the minimum poling domain length of $\sim 1$ $\mu$m, which may pose a challenge with current fabrication technologies. 
The issue may be overcome by using a longer crystal length ($L > 5$ mm), which can have a sufficient number of unit poling domains with a lower value of $\beta$. 
Furthermore, given the rapid advances in the poling technology, submicron poling may also become commercially available in the near future.

%%%%%%%%%%%%%%%%%%%%%%%%%%%%%%%%%%%%%%%%%%%%%%%%%%%%

\section{Heralded single-photon purity versus spectral range}

\begin{figure*}[tbp]
\centering\includegraphics[width=0.8\textwidth]{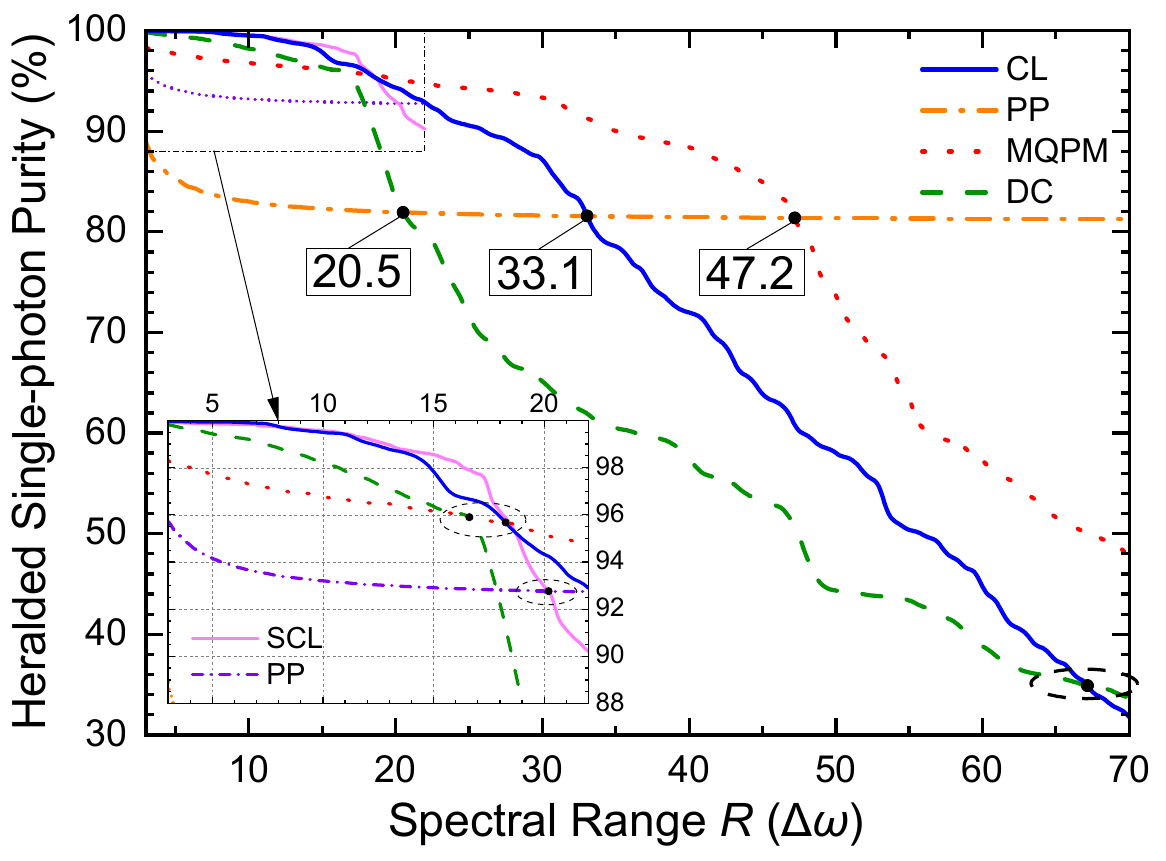}
\caption{Heralded single-photon purity versus spectral range for different poling schemes applied to the GVM case (\romannumeral1)  ($\lambda_i$ = 1550 nm). The spectral range $R$ is shown in the unit of $\Delta \omega$. 
The inset shows the purity for the spectral range up to $\sim 20 \Delta \omega$. 
The purity for the SCL and PP schemes applied to the GVM case (\romannumeral2) ($\theta=2^{\circ}$) is also plotted in the inset.
}
\label{purity-FWHM}
\end{figure*}
In the previous section, we calculated the purity of heralded single photons with $R = 10 \Delta \omega$. Here, we investigate the purity for further extended spectral ranges, which can simulate collection bandwidths in a practical optical setup. We take the GVM case (\romannumeral1) for the investigation. 
Figure \ref{wide range} (a,b) shows the JSA $\left| f\left( \omega _s,\omega_i \right) \right|$ optimized by the CL scheme and its frequency anti-correlation distribution $| f ( \omega_s, \omega_{s0} + \omega_{i0} -\omega_s )|$, which clearly illustrates the peripheral lobes and the noise spectrum. For comparison, we also show results obtained by other poling schemes: Figure \ref{wide range} (c,d) shows the result with the PP structure, and (e,f) and (g,h) are obtained by poling optimization schemes using spatially controlled multi-order quasi-phase-matching conditions (MQPM) \cite{Branczyk2011, Kaneda2021} and poling duty cycles (DC) \cite{Dixon2013, Chen2017, Cui2019, Chen2019, Cai2022}, respectively (poling designs produced by the MQPM and DC schemes are shown in Appendix B). 
The JSA shown in Fig. \ref{wide range} (a,c,e,g) is mapped with $R = 10\Delta \omega$ (which is the same as Fig. \ref{GVM}), while the plot range of the anti-diagonal distributions in Fig. \ref{wide range} (b,d,f,h) is $R = 70\Delta \omega$. 
We find that although the JSA with our employed CL scheme has less noise and higher purity than the other schemes for $R = 10 \Delta \omega$, the noise spectrum outside the range is comparable to or larger than those of the other schemes. 
A similar observation for PMFs at the telecom L-band was reported in \cite{Graffitti2018}. 
A recent review shows the definition of different poling schemes and introduces experimental progress \cite{Weiss2025}.

Figure \ref{purity-FWHM} shows the spectral purity versus spectral range, which can illustrate the impact of noise on broadband photon collection. 
We see that more advanced and complex poling schemes using more various patterns of poling structures can produce higher spectral purity for $R \lesssim 18 \Delta \omega$. 
However, the simplest PP scheme has the lowest degradation in purity as $R$, superior to the DC, CL, and MQPM schemes when the spectral range exceeds $R = 20.5 \Delta \omega$, $33.1 \Delta \omega$, and $47.2 \Delta \omega$, respectively. 
Thus, although the advanced poling optimization schemes are intended to produce high-purity heralded single photons without \textit{tight} spectral filtering ($\lesssim \Delta \omega$), one needs \textit{loose} spectral filtering ($\sim 10 \Delta \omega$) to remove the noise photons with minimum degradation of spectral heralding efficiency, i.e., the probability of a filtered signal photon conditional on the presence of a filtered idler photon. 
For example, for $R = 10 \Delta\omega$, $P_O = 99.5$\% and $>$ 99\% spectral heralding efficiency are simultaneously achieved.
A similar investigation was also performed for the GVM case (\romannumeral2) ($\theta =2^{\circ}$) using the SCL and PP schemes. 
As observed in the GVM case (\romannumeral1), the purity for the SCL scheme applied to the GVM case (\romannumeral2) is also largely degraded as $R$ and only superior to the PP scheme for $R < 20.2 \Delta \omega$.

%%%%%%%%%%%%%%%%%%%%%%%%%%%%%%%%%%%%%%%%%%%%%%%%%%%%

\section{Conclusion}
In this work, we have demonstrated the numerical optimization of spectrally pure heralded-single photon sources at the telecom O-band, where pure heralded single photons can be transmitted through standard telecom optical fiber cables with low loss and low group-velocity dispersion. 
The spectral purity of heralded single photons is optimized by appropriate Type-II GVM conditions and poling structures in a KTP crystal. After determining the range of pump and idler wavelengths yielding positive GVM angles, the CL and SCL schemes were applied to find optimal poling structures and corresponding PMFs on several GVM cases. The JSA constructed by each optimized PMF achieved the heralded single-photon purity of $>99$\%. One of the GVM cases has the pump, signal, and idler wavelengths of 710 nm, 1310 nm, and 1550 nm, which is feasible with current off-the-shelf laser and detector technologies.
The other cases may be implemented by utilizing state-of-the-art technologies of single-photon detectors operated at the wavelength of $> 1.9$ $\mu$m and/or $\sim1$ $\mu$m poling domain fabrication techniques.
Note, however, that the minimum poling domain length can be increased by using a longer crystal length than that of our simulation ($L = 5$ mm). All of the GVM cases are compatible with commercially available pump laser sources such as Ti:sapphire lasers and frequency-doubled optical parametric oscillators, demonstrating the feasibility of our optimization schemes thoroughly.
Moreover, we investigated the noise photon spectra of our optimized SPDC source and those optimized by other poling optimization schemes. 
It shows that an advanced poling optimization scheme using more various poling patterns has less noise and higher spectral purity only in a certain spectral range, and appropriate bandpass filtering is necessary to achieve high purity practically. 
These observations provide insights into the design of an entire optical system of the optimized heralded single-photon sources at the telecom O-band, from the pump pulse preparation and the detection of SPDC photons.
Our approach can also help optimize the source in other wavelength ranges and nonlinear crystals. As shown in Appendix C, the optimization procedure was successfully applied to the source in the telecom C-band. We anticipate that our research will contribute to developing and benchmarking heralded single-photon sources and photonic quantum information applications at the telecom O-band.

%\section*{Funding.}
%This work is supported by JSPS KAKENHI (Grant Nos. JP21K18902, JP22H01965, and JP24K21520), JST ERATO (Grant No. JPMJER2402), and JST PRESTO (Grant No. JPMJPR2106). 
%R. -B. Jin is supported by the National Natural Science Foundation of China (Grant Nos. 92365106 and 12074299).
%W. -H. Cai is supported by JST SPRING (Grant No. JPMJSP2114).

\section*{Acknowledgments.}
We thank Dr. Francesco Graffitti and Prof. Keiichi Edamatsu for the helpful discussion. 
This work is supported by JSPS KAKENHI (Grant Nos. JP21K18902, JP22H01965, and JP24K21520), JST ERATO (Grant No. JPMJER2402), and JST PRESTO (Grant No. JPMJPR2106). 
R. -B. Jin is supported by the National Natural Science Foundation of China (Grant Nos. 92365106 and 12074299).
W. -H. Cai is supported by JST SPRING (Grant No. JPMJSP2114).

%\section*{Disclosures.}
%The authors declare no conflicts of interest.

%\section*{Data Availability Statement.}  
%Data underlying the results presented in this paper are not publicly available at this time but may be obtained from the authors upon reasonable request.

%\newpage
\section*{Appendix A: Derivation of the target PMF}

\begin{figure}[bp]
\centering\includegraphics[width=0.45\textwidth]
{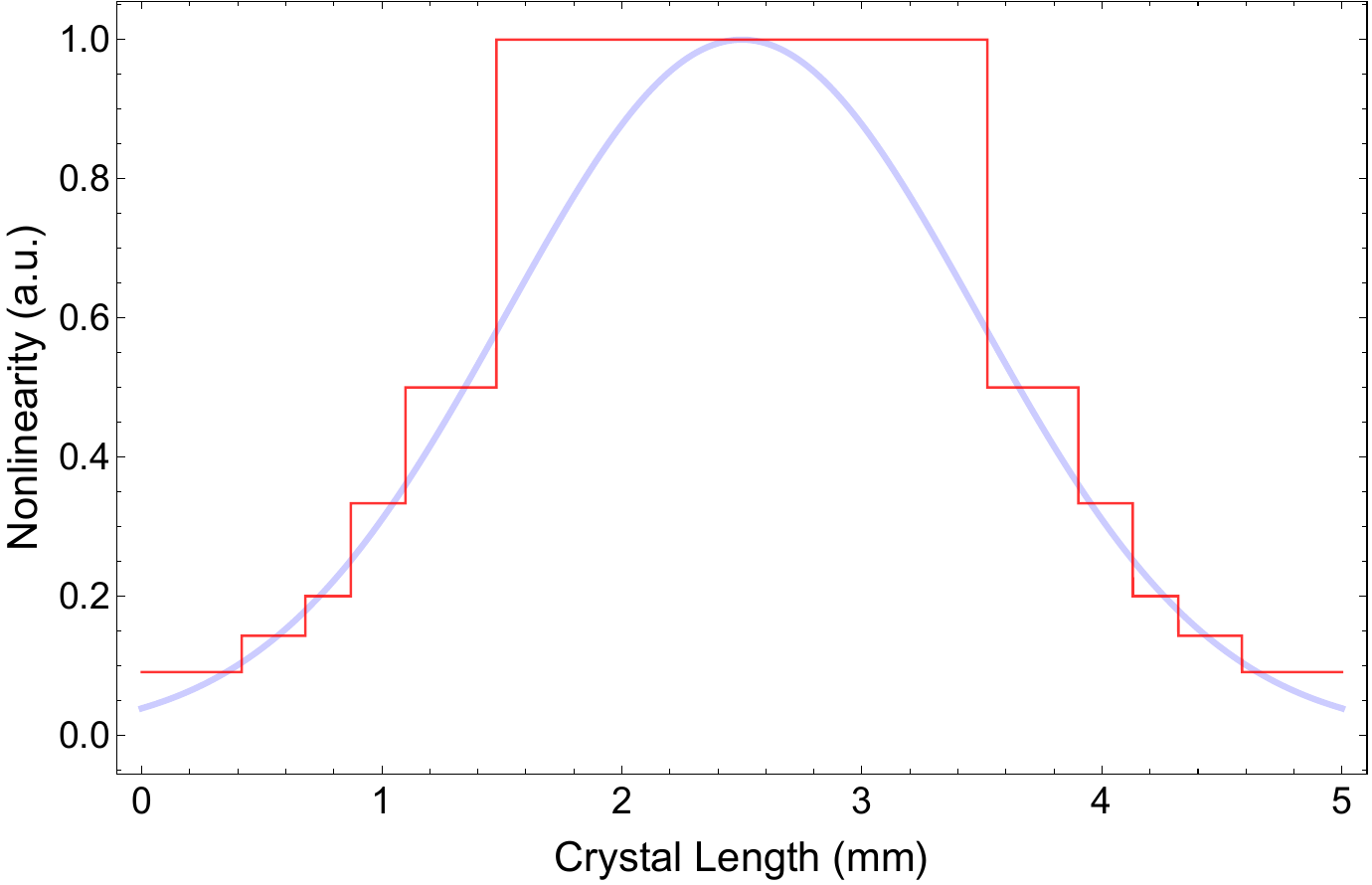}
\caption{
The spatial nonlinearity profile in the KTP crystal obtained by the MQPM scheme (red) and  
the target Gaussian profile (purple). 
}
\label{multi-order}
\end{figure}

\begin{figure}[tp]
\centering\includegraphics[width=0.45\textwidth]{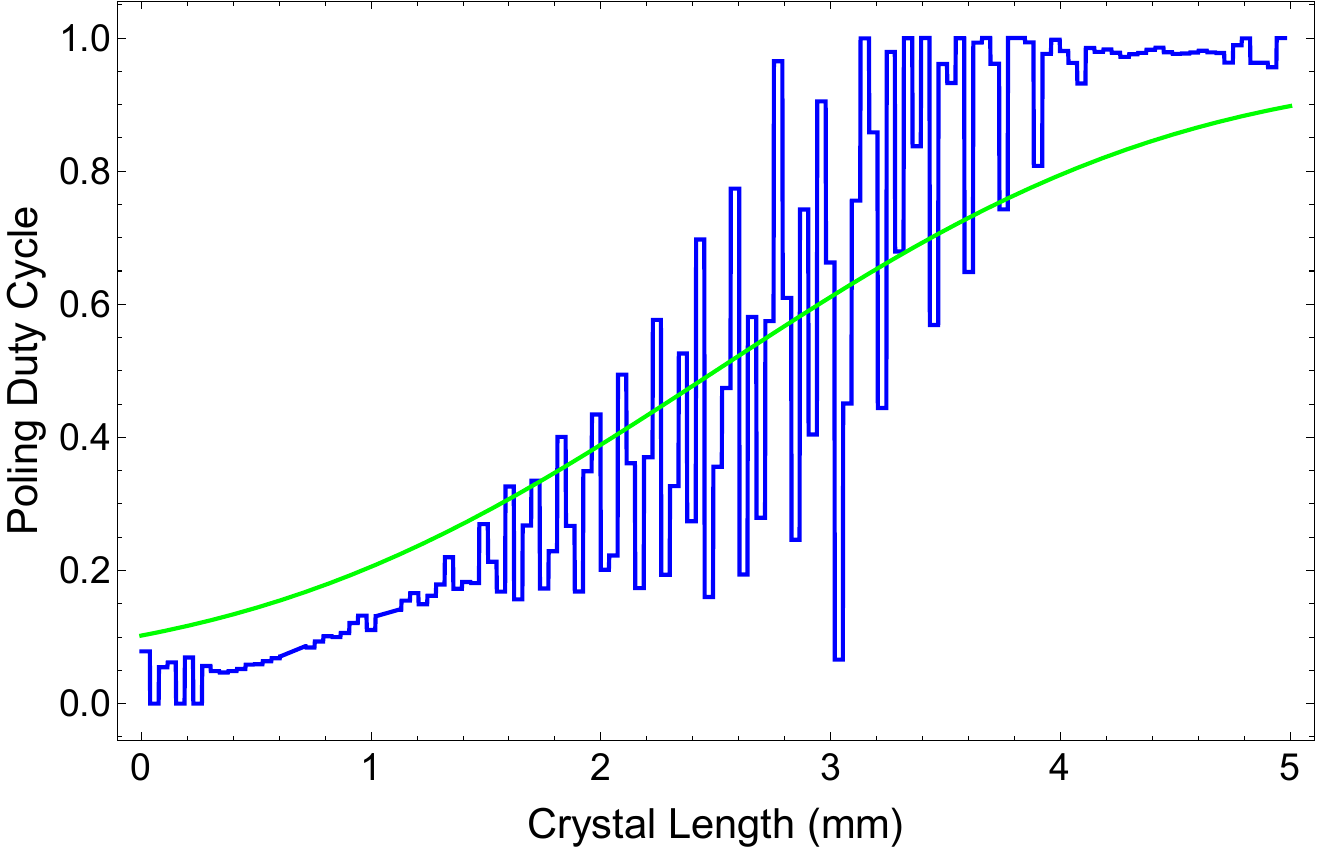}
\caption{
Spatial distribution of the poling duty cycle optimized via the DC modulation scheme (blue) and a Gaussian error function (green), which is used as an initial state of the poling modulation. 
}
\label{duty cycle}
\end{figure}

\begin{table*}[htbp]
\caption{Summary of optimization results for $\lambda_s = $ 1550 nm. The signal photon is $Z$-polarized for the cases (\romannumeral9-\romannumeral12) and $Y$-polarized for the cases (\romannumeral13, \romannumeral14). The purity for the PP and optimized poled schemes are denoted as $P_P$ and $P_O$, respectively.  The optimized pump bandwidth $\Delta\lambda_p$ is used when $P_O$ is calculated.}
\begin{tabular}{ccccccccc}
\hline \hline
Case & $\lambda_y \rightarrow \lambda_z+\lambda_y$ & $\theta$ & $l_c$ ($\mu$m) & $\alpha$ & $\beta$ & $\Delta\lambda_p$ (nm) &$P_P$& $P_O$ \\
\hline
(\romannumeral9)&$ 643.4 \rightarrow 1550 + 1100 $ &  3$^{\circ}$ & 78.95 & 6 & 50 & 7.68 & 91.26\% & 99.62\% \\
(\romannumeral10)&$ 775.0 \rightarrow 1550 + 1550 $ &  40$^{\circ}$ & 22.52 & 4.6 & 4 & 3.36 & 82.34\% & 99.73\% \\
(\romannumeral11)&$ 797.7 \rightarrow 1550 + 1644 $ & 45$^{\circ}$ & 20.94 & 4.8 & 4 & 3.53 & 82.32\% & 99.70\% \\
(\romannumeral12)&$ 971.1 \rightarrow 1550 + 2600 $ &  87$^{\circ}$ & 16.66 & 4.9 & 12 & 14.8 & 91.07\% & 99.35\% \\
\hline
(\romannumeral13)&$ 569.4 \rightarrow  900 + 1550$  & 87$^{\circ}$ & 15.06 & 4.9 & 4 & 3.06 & 91.41\% & 99.77\% \\
(\romannumeral14)&$ 799.2 \rightarrow  1650 + 1550$ & 45$^{\circ}$ & 24.37 & 4 & 5.5 & 4.27 & 82.32\% & 99.77\% \\
\hline \hline
\end{tabular}\\
\label{result-1550}
\end{table*}

The target Gaussian PMF $\phi_{\textrm{T}}(\Delta k_0, z)$ in our optimization procedure is derived as follows. 
We first define the ideal Gaussian PMF $\phi_{\textrm{I}}( \Delta k)$ with 
\begin{equation}\label{phi_I_k}
\phi_{\textrm{I}}(\Delta k) = \exp \left[-\frac{1}{2}\left(\Delta k-\Delta k_0\right)^{2}\sigma^{2}\right],
\end{equation}
where $\sigma$ is the spatial width of a Gaussian nonlinearity profile and $\Delta k_0$ is the phase mismatch at the center frequencies of pump, signal, and idler modes.
By inverse Fourier transform of $\phi_{\textrm{I}}( \Delta k)$, the ideal spatial nonlinearity $g_{\textrm{I}}(z)$ can be obtained: 
\begin{equation}\label{g_I_z}
\begin{aligned}
g_{\textrm{I}}(z) & = \frac{1}{\sqrt{2\pi}} \int_{-\infty}^{\infty} \phi_{\textrm{I}}(\Delta k) e^{-i\Delta kz}d(\Delta k) 
\\ & = \frac{1}{\sigma}\exp(-\frac{z^2}{2\sigma^2}-i\Delta k_{0}z).
\end{aligned}
\end{equation}
Since $g_{\textrm{I}}(z)$ is distributed over an infinite length, the reproduction of $g_{\textrm{I}}(z)$ is impossible with a nonlinear crystal having a finite length. 
Therefore, the target PMF is constructed by integrating $g_{\textrm{I}}(z')$ over a (partial) region of the nonlinear crystal ($z\in [-L/2, L/2]$):
\begin{equation}\label{phi_I_3D}
\begin{aligned}
\phi_{\textrm{T}}(\Delta k,z) 
= & \frac{1}{\sqrt{2\pi}} \int_{-L/2}^{z} g_{\textrm{I}}(z') e^{i\Delta kz'}dz' 
\\ = & \frac{\mathcal{N}}{2} e^{-\frac{1}{2}\left(\Delta k-\Delta k_0\right)^2 \sigma^2} \biggl[\operatorname{erf}\left(\frac{z-i\left(\Delta k-\Delta k_0\right) \sigma^2}{\sqrt{2} \sigma}\right) 
\\ + & \operatorname{erf}\left(\frac{L+2i\left(\Delta k-\Delta k_0\right) \sigma^2}{2\sqrt{2} \sigma}\right)\biggr],
\end{aligned}
\end{equation}
where $\mathcal{N}$ is scaling factor, and $\operatorname{erf}()$ is the error function. 
Due to the satisfaction of the quasi-phase-matching condition (i.e., $\Delta k=\Delta k_0$) in the CL and SCL schemes, $\phi_{\textrm{T}}(\Delta k,z)$ is simplified as
\begin{equation}\label{phi_T_z_origin}
\begin{aligned}
\phi_{\textrm{T}}(\Delta k & = \Delta k_0,z\to z-L/2) 
\\ & = \frac{\mathcal{N}}{2}\left[\operatorname{erf}\left(\frac{L}{2\sqrt{2} \sigma}\right)+\operatorname{erf}\left(\frac{z-L/2}{\sqrt{2} \sigma}\right)\right].
\end{aligned}
\end{equation}
%\begin{equation}\label{scaling}
%{\mathcal{N}}=2\sqrt{\frac{2}{\pi}}\sigma.
%\end{equation}
Here, ${\mathcal{N}}=\sigma\sqrt{8/\pi}$, and the crystal position is shifted ($z\to z-L/2 $) for computational simplicity in the poling optimization. 

\section*{Appendix B: Details of other poling schemes}\label{Appendix-b}
In Sec. 4, we compared the performance of our employed CL and SCL schemes with the PP, MQPM, and DC schemes. Here, we show the details of poling structures optimized by the MQPM and DC schemes.  

The MQPM scheme applies different orders of quasi-phase-matching conditions to control effective nonlinearity. In general, the effective nonlinearity is inversely proportional to the order of quasi-phase matching $m$. 
In this scheme, to achieve approximate Gaussian nonlinearity, the lowest-order ($m =1$) quasi-phase-matching region is placed at the middle of the crystal, and $m$ is increased as the distance from the center of the crystal \cite{Branczyk2011, Kaneda2021}. 
The optimized nonlinearity profile by the MPQM scheme is shown in Fig. \ref{multi-order}. We used six different poling periods satisfying 1st to 11th-order quasi-phase-matching modulations. 
The scheme offers the advantage of simplified fabrication and can be seen as a special case of the CL scheme, although its attainable purity is less than that of the CL scheme, as shown in Fig. 7 in the main text.

%\subsection*{B.2. Duty-cycle (DC) modulation}

In the DC scheme, the effective nonlinearity produced within a fixed poling period of $2 l_c$ is controlled by the duty cycle of UP and DOWN poling domain lengths \cite{Dixon2013, Chen2017, Cui2019, Chen2019, Cai2022}.
We employed the advanced scheme \cite{Cui2019, Cai2022} utilizing a particle swarm optimization algorithm to maximize the purity.
Figure \ref{duty cycle} shows our optimized spatial distribution of duty cycles.
As shown in Fig. \ref{wide range}(h), the JSA is produced in an asymmetric form, which may result in a large reduction in purity as $R$; this will be investigated in future work. 
Moreover, the DC scheme needs shorter poling domains than the other methods and may be infeasible with current technology.

\section*{Appendix C: Optimized heralded single-photon sources at 1550 nm}\label{Appendix-c}
We apply our optimization procedure for heralded single-photon sources at 1550 nm in the telecom C-band. 
Table \ref{result-1550} shows the summary of the optimization results for the GVM cases (\romannumeral9-\romannumeral14) in Fig. \ref{GVM-theta}.   
All selected GVM cases achieved the purity $P_O > 99\%$.
The GVM cases (\romannumeral10) and (\romannumeral13) are close to experimental works \cite{Kaneda2016, Chen2019, Graffitti2018optica} demonstrated with various poling schemes. 
The estimated spectral purity $P_O$ is superior to or comparable with the previous works.

%\bibliography{CPKTP}% Produces the bibliography via BibTeX.
%

\end{document}